\documentclass[aps,showpacs,prb,twocolumn]{revtex4}
\usepackage{times}
\usepackage{graphicx}
\begin{document}
\title{Kohn anomalies and non-adiabaticity in doped carbon nanotubes}
\author{Nicolas Caudal}
\author{A. Marco Saitta}
\email{saitta@impmc.jussieu.fr}
\author{Michele Lazzeri}
\author{Francesco Mauri}
\affiliation{Institut de Min\'eralogie et Physique des Milieux
Condens\'es, CNRS-UMR 7590, Universit\'e Pierre et Marie
Curie-Paris 6, Universit\'e Denis Diderot-Paris 7, IPGP, F-75252
Paris, France}
\date{\today}
\begin{abstract}

The high-frequency  Raman-active phonon modes of metallic single-walled carbon
nanotubes (SWNTs) are thought to be characterized by Kohn
anomalies (KAs) resulting from the combination of SWNTs intrinsic
one-dimensional nature and a significant electron-phonon coupling
(EPC). KAs are expected to be modified by the doping-induced tuning
of the Fermi energy level $\epsilon_F$, obtained through the
intercalation of SWNTs with alkali atoms or by the application of a
gate potential. We present a Density-Functional Theory (DFT) study
of the phonon properties of a (9,9) metallic SWNT as a
function of electronic doping. For such study, we
use, as in standard DFT calculations of
vibrational properties, the Born-Oppenheimer
(BO) approximation.
We also develop an analytical model
capable of reproducing and interpreting our DFT results. Both DFT
calculations and this model predict, for increasing doping levels,
a series of EPC-induced KAs in the vibrational mode parallel to
the tube axis at the $\mathbf\Gamma$ point of the Brillouin zone,
usually indicated in Raman spectroscopy as the $G^-$ peak. Such
KAs would arise each time a new conduction band is populated.
However, we show that they are an artifact of the
BO approximation.
The inclusion of non-adiabatic (NA)
effects dramatically affects the results, predicting KAs at
$\mathbf\Gamma$ only when $\epsilon_F$ is close to a band crossing
$E_{X}$. For each band crossing a double KA occurs for
$\epsilon_F=E_{X}\pm \hbar\omega/2$, where $\hbar\omega$ is the
phonon energy. In particular, for a 1.2 $nm$ metallic nanotube, we
predict a KA to occur in the so-called $G^-$ peak at a doping
level of about $N_{el}/C=\pm 0.0015$ atom ($\epsilon_F\approx \pm
0.1 ~eV$) and, possibly, close to the saturation doping level
($N_{el}$/$C$$\sim$ 0.125), where an interlayer band crosses the
$\pi^*$ nanotube bands. Furthermore, we predict that the Raman
linewidth of the $G^-$ peak significantly decreases for
$|\epsilon_F| \geq \hbar\omega/2$. Thus our results provide a tool
to determine experimentally the doping level from the value of the
KA-induced frequency shift and from the linewidth of the $G^-$
peak. Finally, we predict KAs to occur in phonons with finite
momentum \textbf{q} not only in proximity of a band crossing, but
also each time a new band is populated. Such KAs should be
observable in the double-resonant Raman peaks, such as the
defect-activated $D$ and peak, and the second-order peaks $2D$ and
$2G$.

\end{abstract}

\maketitle

\section{Introduction}

Since their discovery in 1991~\cite{Iijima91}, carbon nanotubes
have raised an enormous interest both from the academic and the
technological points of view. They exhibit in fact a variety of
exciting features: their quasi-one-dimensional nature, due to a
diameter of 1-2 $nm$ and a length of up to several micrometers,
makes them sharp probes for scanning tunneling microscopes and an
excellent model for one-dimensional physics. Their mechanical and
tensile strength make them of great interest in composite
materials, and, most significantly of all, they have very unusual
and extremely promising electronic properties, displaying metallic
or semiconducting behavior according to their structure and
helicity~\cite{Saito,Reich}. The electronic properties of SWNTs,
already particularly interesting from the technological point of
view, promise to be the future of nano-electronics due to their
\emph{tunability}, achieved by doping the nanotubes through
intercalation with alkali atoms
\cite{Ye03,Bendiab01,Bendiab01b,Bendiab01c,BendiabThesis,Furtado05,Meunier02,Liu03,Rauf04,Bantignies05}
or application of a gate
potential~\cite{Corio04,Cronin04,Rafailov05,Wang06}. However,
there are still a number of experimental challenges to be solved
in order to fully develop SWNT-based nano-electronic technology,
in particular the low-cost industrial-scale synthesis of nanotubes
of good chemical purity, crystalline quality and given helicity.

An experimental tool largely used to characterize SWNTs is Raman
spectroscopy
\cite{Bendiab01,Bendiab01b,Bendiab01c,BendiabThesis,Ye03,Furtado05,
Maultzsch02,Maultzsch03,Cronin04,Rafailov05,Maultzsch05,Son06,Wang06,Jorio02,Jorio05}.
Typical Raman spectra of carbon nanotubes display a peak around
150-300 $cm^{-1}$, due to the radial breathing mode (RBM), which
has been recently used as a tool to infer the size and chirality
of the nanotubes~\cite{Maultzsch05,Jorio05}. Other important
features of SWNT Raman spectra include a peak around 1350
$cm^{-1}$, activated by defects and impurities, and known in the
literature as $D$ peak, and a large structure around 1570
$cm^{-1}$, due to modes tangential to the nanotube and known as
$G$ peak. This last feature is thought to have two components,
usually referred to as $G^+$ and $G^-$, originating from the
$E_{2g}$ in-plane modes of graphite. In
refs.~\cite{Lazzeri06,Piscanec06,Popov06} it has been shown that
in metallic SWNTs the $G^+$ component corresponds to the
tangential vibrational mode \emph{perpendicular} to the nanotube
axial direction, while the $G^-$ component corresponds to the
tangential vibrational mode \emph{parallel} to the nanotube axial
direction. In the following, we will refer to the former as ``the
nanotube TO tangential mode'', and to the latter as ``the nanotube
LO axial mode''. Some Raman
studies~\cite{Bendiab01,Bendiab01b,Bendiab01c,BendiabThesis} show
that the frequency of the $G$ peak increases up to 1600 $cm^{-1}$
at low doping levels, and then suddenly drops to about 1550
$cm^{-1}$ at the saturation threshold of alkali intercalation,
estimated around a number of electrons per carbon atom
$N_{el}/C=0.12$ (MC$_{\rm 8}$).

Other experimental~\cite{Maultzsch02,Rafailov05,Furtado05} and
theoretical~\cite{Dubay02,Dubay03,Akdim05} works on the effect of
doping on SWNTs report a similar $G$-peak softening or even a
Luttinger-Fermi liquid transition~\cite{Rauf04}. Since the LO
axial and TO tangential modes are particularly sensitive to the
electronic structure of SWNTs around the Fermi energy
~\cite{Dubay02,bohnen04,connetable05,Lazzeri06,Popov06}
$\epsilon_F$, and $\epsilon_F$ directly depends on the charge
doping level, a profound understanding of the interplay between
the vibrational and the electronic properties of nanotubes looks
to be crucial for technological development.

In this work we report our theoretical study of the electronic and
vibrational properties of doped SWNTs, based on DFT
first-principles calculations and analytical results. We will show
in the following that: \emph{i)} the vibrational properties of
SWNTs can be obtained from the so-called \emph{electronic
zone-folding} of a graphene sheet; \emph{ii)} their behavior
as a function of the charge doping level can be determined by the
knowledge of electron-phonon coupling (EPC) in graphene;
\emph{iii)} ordinary quantum-mechanics calculations
relying on the adiabatic Born-Oppenheimer approximation
fail when applied to SWNTs, where non-adiabatic effects are
enhanced by their intrinsic one-dimensional nature.

Our paper is organized as follows: in section~\ref{theo} we will
describe our theoretical framework, and in particular how the
Raman-active modes of a nanotube can be accurately obtained
through an appropriate electronic sampling of the graphene
Brillouin Zone (BZ). We will then show in section~\ref{model} that
the DFT results can be almost perfectly reproduced by an integral
model, that uses the graphite EPC and the slope of the electronic
bands around the (undoped) Fermi level $\epsilon^0_F$ (usually
referred to as the $\pi$ and $\pi^*$ bands) as external inputs,
and that becomes analytical in the limit of vanishing temperature.
The results on the DFT and the model-derived LO axial and TO
tangential modes of a SWNT will be presented, showing that
Kohn anomalies (KA)
would occur, within the adiabatic approximation, each time a new
electronic band is populated by the electrons at $\epsilon_F$. In
section~\ref{nonadia} we show that when the Born-Oppenheimer
approximation is lifted the outcome is dramatically different, and
that a drop in the frequency of the $G^-$ peak occurs at a doping
level such that $\epsilon_F-\epsilon^0_F\approx 0.1 ~eV$ or close
to the saturation level. Section~\ref{conclusions} will be devoted
to the discussion of the physical properties of metallic SWNTs
that are experimentally accessible, and to the conclusions.

\section{Theoretical background}\label{theo}

\subsection{General description of graphene and SWNTs}\label{graphene}

Graphene is a semimetal, and its highest valence and lowest
conduction bands have a circular conical surface shape
around the points \textbf{K} and \textbf{K$^\prime$} of the
hexagonal first Brillouin zone at zero doping; its Fermi surface is reduced to
the vertexes of the cones at those two points.

The structure of a SWNT is obtained by rolling a sheet of 2D
graphene, and closing both ends with fullerene-like semi-spheres.
However, nanotubes are commonly studied as infinite 1D crystals,
{\it i.e.} neglecting the details of those ends. Nanotubes are
then uniquely determined by the knowledge of the chiral vector
${\bf C_h}=n{\bf a_1}+m{\bf a_2}$, where ${\bf a_1}$ and ${\bf a_2}$
are the basis vectors of graphene, $n$ and $m$ are integers~\cite{Saito,Reich}.
The translational vector
determines the periodicity along the tube axis and is
${\bf T}=\frac{(2m+n)}{d_R}{\bf a_1}-\frac{(2n+m)}{d_R}{\bf a_2}$,
where $d_R$ is the greatest common divisor of $(2m+n)$ and $(2n+m)$.
Nanotubes with indexes such that $n-m=3k$,
where $k$ is an integer, are metallic, otherwise they are
semiconductors. Nanotubes with identical indexes, \emph{i.e.
n=m}, are thus always metallic, and are usually called ``armchair''
tube.

Electronic zone folding (EZF) is a common approximation for the electronic band structure
of a SWNT, which consists in neglecting the curvature of the SWNT.
Within the EZF the electronic states of a SWNT are approximated
by the electronic states of a graphene sheet
having wavevector ${\bf k}$ such that ${\bf k}\cdot{\bf C_h}=2\pi\nu$, being
$\nu$ an integer.
The wave-vectors allowed for the SWNT satisfy
\begin{equation}\label{nano_nu}
\mathbf{k}=\nu \mathbf{k}_\perp + k \mathbf{k}_\parallel / \|
\mathbf{k}_\parallel \|
\end{equation}
where $k$ is a real, $\mathbf{k_\parallel}$ is parallel to the tube axis,
$\mathbf{k_\perp}$ is perpendicular to the tube axis and satisfies
$\|\mathbf{k}_\perp\|=2 \pi/ \|\mathbf{C}_\mathrm{h}\|=2/d$, $d$
being the tube diameter.

This way of constructing  the electronic band structure accounts for
the fact that SWNTs can be either metallic or semiconducting.
In fact, if in SWNTs a line of allowed wave vectors \textbf{k}
crosses the \textbf{K} or the \textbf{K$^\prime$} point, the
nanotube is metallic, otherwise it is a semiconductor. In
particular, in the BZ of armchair metallic SWNTs the section of the
vertical plane containing the $\nu=0$ line of \textbf{k} vectors
and the conical bands consists of two straight lines of slope
$\pm\beta$, while the conic sections for $|\nu|\ge 1$ consist of
arms of hyperbolae. The energy difference between the maximum of
the lower arm and the minimum of the upper arm are usually
indicated as $E_{\nu\nu}^M$ in the metallic case (see also
Fig.~\ref{fermi}).

Previous works~\cite{Saito,Reich,Zolyomi04,Piscanec06,Lazzeri06}
have shown that the effects of the nanotube curvature can be
neglected as a first approximation for SWNTs whose diameter is
larger than 1 $nm$ (the most common nanotubes present in
experimental samples) and the electronic properties of a nanotube
can be quite accurately obtained from EZF.

\subsection{Electronic zone-folding for phonon calculations}

The explicit calculation of the full dynamical matrix of an
infinite nanotube, even within an efficient scheme such as
Density-Functional Perturbation Theory (DFPT)~\cite{Baroni01}, is
a quite demanding task from the computational point of view. On
the other hand, the SWNT vibrational modes typically observed in
Raman spectra can be traced back to the phonon and elastic modes
of an isolated graphene sheet.

However, it has been shown that the phonon dispersions of SWNTs
cannot be as accurately reconstructed from the graphene ones as
the electronic bands, because of the presence of KAs which behave
differently in graphene~\cite{Piscanec04} and metallic SWNTs
~\cite{Dubay02,bohnen04,connetable05,Lazzeri06,Piscanec06,Popov06}.
The KAs are determined by singularities in the electron screening
and such singularities depends on the dimensionality of the
electron Brillouin zone. The dimensionality is two for graphene
and one for SWNTs. This reduced SWNT dimensionality is due to the
quantization ({\it confinement}) of the electronic wavevector
around the nanotube circumference, which is actually described by
the EZF. This last consideration suggests a practical scheme for
the calculation of the phonon dispersion in SWNTs which neglects
curvature, but fully takes into account the more important
confinement effects. Phonons in nanotubes are well approximated by
the phonons of a flat graphene sheet, if the calculation is done
performing the electronic Brillouin-zone integration on the lines
of the electronic zone-folding (Eq.~\ref{nano_nu}). Such phonon
calculation method (phonon-EZF) was introduced in
refs.~\cite{Lazzeri06,Piscanec06}. A phonon-EZF calculation
requires the use of a unit cell containing two atoms and thus, is
clearly much less computationally demanding than full calculations
on an actual SWNT, which requires a unit-cell with tens or
hundreds of atoms. In Subsect.~\ref{accuracy} we will demonstrate
that, for the preset study, curvature effects are negligible and
the phonon-EZF provides an accurate description of the
high-frequency phonon modes.

\subsection{Computational details}
Our first-principles calculations are based on DFT, within the
plane-wave (PW)/pseudopotential scheme implemented in the
Quantum-ESPRESSO code~\cite{PWSCF}. We adopt a
Perdew-Burke-Erzherhof gradient corrected functional, and an
ultrasoft pseudopotential to describe the $C$ atom. A PW kinetic
energy cutoff of 30 Ry is sufficient to ensure convergence on the
structural, electronic, and vibrational properties. We choose as a
case-study the (9,9) metallic armchair SWNT, containing 36
distinct carbon atoms, and having a tube diameter of 1.24 $nm$.
The full nanotube calculations are performed by setting the
nanotube in an infinite lattice of hexagonal symmetry in the plane
perpendicular to the tube axis, mimicking thus the bulk effect of
a real nanotube bundle or an isolated tube by tuning the lattice
parameter between 1.56 $nm$ and 1.72 $nm$. The $c$ axis is
maintained constant. Integrations in the nanotube one-dimensional
Brillouin zone have been performed by using regular grids of
\textbf{k} points along the $c$ reciprocal axis. We use a
Fermi-Dirac electronic smearing of 0.01 Ry (corresponding to an
electronic temperature of 1578 K) for SWNT full phonon
calculations, and of 0.002 Ry (315 K) for structural calculations
and frozen-phonon tests (see below). Grids of 8 and 40 \textbf{k}
points were sufficient, respectively, to ensure a good convergence
($\sim$ 5 $cm^{-1}$) on the LO axial frequencies in the two cases.

In the phonon-EZF calculation we use a graphene
hexagonal unit cell of lattice
parameter $a_{\mathrm{exp}}=2.47$ \AA{} and $c=5$ \AA. The number
of points is chosen as to ensure phonon frequencies converged
within about 0.3 $cm^{-1}$, and is inversely proportional to
temperature. At a temperature of 315 K, the total number of points
is 3240, equivalent to 180 points along the $c$ direction of the
reciprocal space of the (9,9) nanotube, while at 1578 K this
number reduces to 720, equivalent to 40 points in the nanotube
case.
For both the full nanotube and the phonon-EZF case, the effect
of (low) charge doping
is simulated by adding an excess electronic charge which is compensated
by a uniformly charged backgournd. This is done using the
standard implementation of the ESPRESSO-package~\cite{PWSCF}.
This approach is known
to describe very well the properties of doped
SWNTs~\cite{Margine06}. This is justified even in the case of
alkali intercalation by the report that at low doping levels, {\it
i.e.} up to MC$_{\rm 15}$, the electronic charge of the metal
atoms is thought to be completely transferred~\cite{Lu04}, and
that the effect of the intercalation is essentially described by
taking into account the sole effect of the charge transfer.

\begin{figure}
\centerline{\includegraphics[width=8.25cm]{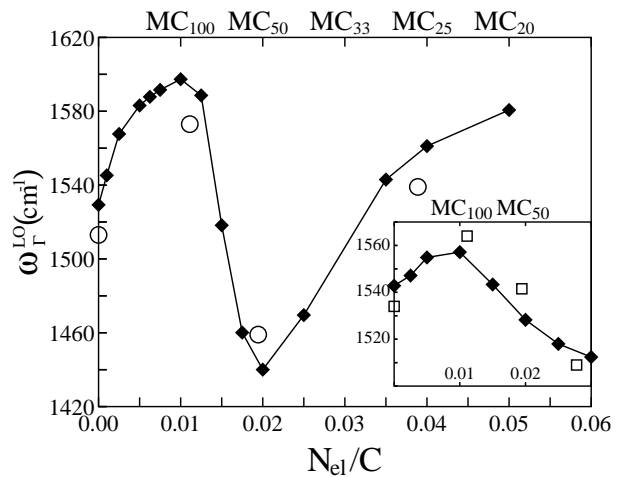}}
\caption{\label{compare} Variation of the LO axial frequencies
with electron doping at 315 K, calculated from phonon-EZF
(full diamonds) or through full-nanotube
frozen-phonon (open circles). The solid line is a guide to the
eye. The metal/carbon composition corresponding to the doping
level is indicated on the top of the figure. Inset: variation with
electronic doping of the phonon-EZF LO axial frequency (full diamonds)
compared to a SWNT full DFPT dynamical matrix calculation (open
squares), at 1578 K.}
\end{figure}

\subsection{Accuracy of the method of phonon calculation}
\label{accuracy}

The first step of our study is to assess the accuracy of the
phonon-EZF method with respect to the SWNT
phonon calculations. In Fig.~\ref{compare} we show the
frequency of the LO axial mode as a function of electronic doping
at 315 K, obtained both through the phonon-EZF method or by an explicit
nanotube frozen-phonon calculation. In the latter case, the atoms
are displaced along the axial direction of the nanotube, and the
frequency is obtained by the curvature of the displacement energy.
The convergence of this single-mode frozen-phonon frequency has
been confirmed by testing the result with a more accurate sampling
of the BZ, that is up to 200 uniformly-spaced \textbf{k} points in
the one-dimensional BZ grid. In the inset we analogously compare
the phonon-EZF LO axial frequency with a full DFPT nanotube phonon
calculation at 1578 K. In this latter case, all the 108 $\Gamma$
vibrational modes of the nanotube are explicitly calculated; a
lower temperature, and thus a correspondingly larger grid of at
least 40 \textbf{k} points, would be computationally too demanding
for a complete and accurate study of the problem. Both graphs show
that phonon-EZF not only determines the nanotube phonon frequencies in
good agreement with explicit SWNT calculations, but it also
captures very well the strikingly non-monotonic dependence of the
LO axial mode with respect to electronic doping, to be discussed
later on, confirming that the effects of the SWNT curvature can be
confidently neglected for a (9,9) armchair nanotube, and that phonon-EZF
is a reliable and accurate method.

\begin{figure}
\centerline{\includegraphics[width=8.25cm]{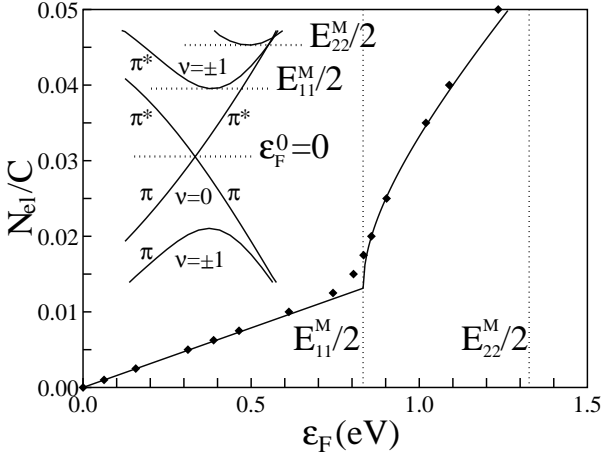}}
\caption{\label{fermi} Variation of the electronic doping per $C$
atom as a function of the Fermi energy $\epsilon_F$. Solid line:
analytical expression~(Eq.\ref{eq:Fermi-dop}) with
$E_{11}^M/2=0.83~eV$; diamonds: DFT
calculations, within phonon-EZF, of the Fermi energy in a (9,9) SWNT at
315 K. The $E_{11}^M/2$ and $E_{22}^M/2$ levels are indicated by
vertical dotted lines. Inset: Calculated electronic bands of the
(9,9) SWNT around the \textbf{K} point. $\epsilon_F^0$ is taken as
the zero energy, $E_{11}^M/2$ and $E_{22}^M/2$ are the minima of
the second and third (hyperbolic) conduction bands. The valence
bands (below $\epsilon_F^0$) and the conduction bands (above
$\epsilon_F^0$) are usually referred to as the $\pi$ and $\pi^*$
bands, respectively. The bands crossing at $\epsilon_F^0$ have the
index $\nu=0$, while the valence and conduction bands immediately
below and above have indexes $\nu=\pm 1$ (see Eq.~\ref{nano_nu}).}
\end{figure}

\subsection{Fermi level in doped SWNTs}\label{subsectionFermi}
The key quantity intervening in the anomalous vibrational
properties of SWNTs is the Fermi energy level $\epsilon_F$. In
undoped graphene and in metallic armchair nanotubes, its value
($\epsilon_F^0$) coincides with the band energy at the crossing of
the $\pi$ and $\pi^*$ bands at the \textbf{K} and
\textbf{K$^\prime$} points of the graphene BZ, and is chosen as
the zero band energy hereafter. Since the results are independent
on the sign of $\epsilon_F$, we consider here and in the following
that $\epsilon_F>0$, without loss of generality. Following the
description of the graphene electronic bands in terms of conic
sections as in subsection~\ref{graphene}, the increase of
$\epsilon_F$ due to electronic doping per carbon atom $N_{el}/C$
in a ($n,n$) SWNT can be analytically expressed, at low doping
levels and a temperature T=0 K, by the following formula:
\begin{equation}\label{fermieq}
N_{\mathrm{el}}/C(\epsilon_F)=\frac{a_0^2\sqrt{3}}{\pi \beta
d}\epsilon_F+ \theta(\epsilon_F - E_{11}^M/2) \frac{2
a_0^2\sqrt{3}}{\pi \beta d} \sqrt{\epsilon_F^2-(E_{11}^M/2)^2},
\label{eq:Fermi-dop}
\end{equation}
where $a_0$ is the lattice parameter, $\beta$ is the slope of the
conical bands and equals 14.1 $eV\frac{a_0}{2\pi}$, and
$\theta(x)$ is the step function. As previously mentioned, the
energies of electronic transitions in SWNTs are usually indicated
as $E_{ii}^{M/S}$,
where, for armchair SWNTs, $i$ coicides with the $\nu$ band index of the
initial and final states, and the superscript refers to metallic
or semiconducting SWNTs. In the case of a ($n,n$) nanotube, assuming
perfect conical bands, $E_{11}^M/2=2\beta/d$. In particular,
for the (9,9) nanotube  $E_{11}^M/2=0.89~eV$.

We report in Fig.~\ref{fermi} the corresponding curve relating
$\epsilon_F$ to $N_{el}/C$, along with our DFT calculations at 315
K. The most significant discrepancy being between the ideal value
of $E_{11}^M/2$ and the calculated one, which is 0.83 $eV$. If
however we treat $E_{11}^M/2$ as an independent parameter, and we
red-shift it by 0.06 $eV$, the agreement is very good. This
indicates that the analytical expressions can be safely used to
develop an integral and analytical model, based on the EPC and the
electronic bands, aimed at describing and predicting the behavior
of the LO axial and TO tangential modes in metallic nanotubes. The
electronic bands of the nanotube, explicitly calculated by DFT,
are shown in the inset around the \textbf{K} point of the BZ,
along with the band nomenclature $\pi/\pi^*,\nu$ used in the
literature and adopted in this work. One can notice that the
minimum of the $\nu=\pm 1$ conduction band is slightly displaced
to the right, and that the bands deviate from the conical shape
far from \textbf{K}. In the following we will use a polynomial fit
of the calculated bands in numerical evaluation of integral
expressions, while the ideal conical shape of the bands
will be used to determine the analytical
limits at low temperature.

\section{Integral and analytical model}\label{model}

\subsection{Electron-phonon coupling contribution to the dynamical
matrix}\label{general}

The frequency of a vibrational mode at a \textbf{q} wavevector, is
obtained from the dynamical matrix $D_{\mathbf{q}}$
\begin{equation}\label{dynmat}
\omega_{\mathbf{q}}= \sqrt{\frac{D_{\mathbf{q}}}{M}},
\end{equation}
where $M$ is the atomic mass of carbon. As shown in
refs.~\cite{Piscanec04,Piscanec06}, the dynamical matrix can be
written as the sum of an EPC direct contribution $\tilde D_{\mathbf{q}}$,
which contains
non-analytical terms giving rise to Kohn anomalies, and a term
containing all the other contributions:
\begin{equation}
D_{\mathbf{q}}= \tilde
D_{\mathbf{q}}+D_{\mathbf{q}}^{\mathrm{other}}
\end{equation}
The direct contribution of EPC to the dynamical matrix can be written as
\begin{equation}
\tilde D_{\mathbf{q}}=\frac{2}{N_{\mathrm{k}}} \sum_{\mathbf{k},
i, f}
\frac{f(\epsilon_{\mathbf{k},i})-f(\epsilon_{\mathbf{k}+\mathbf{q},f})}{\epsilon_{\mathbf{k},
i}- \epsilon_{\mathbf{k}+\mathbf{q}, f}} \cdot
|G_{(\mathbf{k}+\mathbf{q}), f; \mathbf{k}, i}|^2
\end{equation}
where $N_{\mathrm{k}}$ is the number of points in the SWNT
Brillouin zone, $i$ and $f$ are the band indexes indicating the two states
involved in the the electronic transition. Since
we consider only the contribution of the $\pi$ and $\pi^*$ bands,
$i,f=\pi,\pi^*$. The function $f(\epsilon)=\{\exp[(\epsilon-\epsilon_F)/kT]+1\}^{-1}$, is the
Fermi-Dirac distribution;
$G_{(\mathbf{k}+\mathbf{q}), f; \mathbf{k}, i}$
is the EPC matrix element, defined as
\begin{equation}
G_{(\mathbf{k}+\mathbf{q}), f; \mathbf{k}, i}=\langle
\mathbf{k}+\mathbf{q}, f |\Delta V_{\mathbf{q}}| \mathbf{k}, i
\rangle
\end{equation}
where $\Delta V_{\mathbf{q}}$ is the first-order derivative of the
Kohn-Sham self-consistent potential with respect to the atomic
displacements corresponding to a \textbf{q}-vector phonon; $|
\mathbf{k}, i \rangle$ is the Bloch electronic wavefunction.
Hereafter, we develop our model on the basis of the EZF sampling,
and the \textbf{k} vectors are those allowed by Eq.~\ref{nano_nu}.
Since we study phonons close to $\mathbf\Gamma$ we limit ourselves
to wave vectors entirely along $\mathbf{k_\parallel}$ writing
$\mathbf{q}=q \mathbf{k}_\parallel/\| \mathbf{k}_\parallel \|$ and
$\|\mathbf{q}\|\ll\|\mathbf{K}\|$. Therefore $\nu$ is conserved
through the transition $i \rightarrow f$. In the limit of a
nanotube of infinite length we replace $\frac{1}{N_{\mathrm{k}}}
\sum_{\mathbf{k}}$ by $\sum_\nu \frac{T}{2 \pi} \int \mathrm{d}k$,
which yields for the non-analytical part of the dynamical matrix
\begin{equation}\label{eq:start}
\tilde D_{\mathbf{q}}= \frac{T}{\pi} \sum_{\nu, i, f} \int
\mathrm{d}k \frac{f \left[\epsilon_{\nu,
i}(k)\right]-f\left[\epsilon_{\nu,f}(k+q)\right] } {\epsilon_{\nu,
i}(k)-\epsilon_{\nu, f}(k+q)} \cdot |G_{(\mathbf{k}+\mathbf{q}),
f; \mathbf{k}, i}|^2 \label{eq:Dq}
\end{equation}
Neglecting the effects of curvature, the EPC matrix element
$G_{(\mathbf{k}+\mathbf{q}), f; \mathbf{k}, i}$ of a nanotube of
diameter $d$ and longitudinal period $T$ is related to the EPC
$\tilde G_{(\mathbf{k}+\mathbf{q}), f; \mathbf{k}, i}$ of graphene
by the ratio of the unit cells areas (Eq.4 of~\cite{Lazzeri05}):
\begin{equation}
d \pi T |G_{(\mathbf{k}+\mathbf{q}), f; \mathbf{k}, i}|^2
=\frac{a_0^2\sqrt{3}}{2} |\tilde G_{(\mathbf{k}+\mathbf{q}), f;
\mathbf{k}, i}|^2
\end{equation}
The main contribution to the EPC term originates from the
\textbf{K} point of the graphene BZ, so we can define the
wavevector $\mathbf{k'}=\mathbf{k}-\mathbf{K}$. Furthermore, since
KAs originate when the denominator in Eq.~\ref{eq:start} vanishes,
we can restrict the integral in the BZ to a small interval of
width $2\overline{k}$ around \textbf{K} (and \textbf{K'}). We can
also define the angles $\theta$, between \textbf{k$^\prime$} and
\textbf{q}, and $\theta'$, between $\mathbf{k^\prime}+\mathbf{q}$
and \textbf{q}. In refs.~\cite{Piscanec04,Lazzeri06} it was shown
that for phonons close to $\mathbf\Gamma$, the nanotube EPC can be
in principle expressed in terms of the graphene EPC $\langle
G_\mathbf{\Gamma}^2\rangle_\mathrm{F}= 45.60$
($eV$/\AA)$^2$~\cite{Piscanec04}, modulated by a geometric factor:
\begin{equation}
|\tilde G_{(\mathbf{k}+\mathbf{q}), f; \mathbf{k}, i}^{\rm
TO/LO}|^2=\langle G_\mathbf{\Gamma}^2\rangle_\mathrm{F}\left[1\pm
sign(\epsilon_{\nu, f}\cdot\epsilon_{\nu,
i})\cos(\theta+\theta')\right]\label{eq:G}
\end{equation}
where
\begin{equation}
\cos (\theta+\theta')=\frac{k' (k'+q)-(\nu
k_\perp)^2}{\sqrt{k'^2+(\nu k_\perp)^2} \sqrt{(k'+q)^2+(\nu
k_\perp)^2}} \label{eq:cos}
\end{equation}
and the $+$($-$) sign has to be considered for TO tangential(LO
axial) modes. This expression, developed through a first-neighbor
tight-binding model was then quantitatively confirmed by direct
DFT calculations~\cite{Piscanec04}. The TO tangential/LO axial
terms of the non-analytical part of the dynamical matrix can thus
be written as:
\begin{equation}\tilde D_{\mathbf{q}}^{\rm TO/LO}=\frac{a_0^2 \sqrt{3}\langle G_\mathbf{\Gamma}^2
\rangle_\mathrm{F}}{\pi^2 d} \sum_{\nu, f,
i}\int_{-\overline{k}}^{\overline{k}} \mathrm{d}k'
\;\frac{f\left[\epsilon_{\nu, i}(k')\right]- f\left[\epsilon_{\nu,
f}(k'+q)\right]}{\epsilon_{\nu, i}(k')-\epsilon_{\nu, f}(k'+q)}
\cdot \left[1\pm sign(\epsilon_{\nu, f}\cdot\epsilon_{\nu,
i})\cos(\theta+\theta')\right] \label{eq:stat}
\end{equation}
where a factor 2 is included to take into account the contribution
of the two equivalent points \textbf{K} and \textbf{K$^\prime$}.

Since we are interested in phonons close to $\Gamma$, we should
consider the $\mathbf{q}\rightarrow 0$ limit in Eq.~\ref{eq:stat}:
\begin{eqnarray}\label{eq:statgamma}
\tilde D_{\mathbf{\Gamma}}^{\rm TO/LO}&=& \frac{a_0^2
\sqrt{3}\langle G_\mathbf{\Gamma}^2 \rangle_\mathrm{F}}{\pi^2 d}
\int_{-\overline{k}}^{\overline{k}} \mathrm{d}k' \; \left\{
\sum_{\nu, f\ne i}\frac{f\left[\epsilon_{\nu, i}(k')\right]-
f\left[\epsilon_{\nu, f}(k')\right]}{\epsilon_{\nu,
i}(k')-\epsilon_{\nu, f}(k')} \cdot \left[1\pm sign(\epsilon_{\nu,
f}\cdot\epsilon_{\nu, i})\cos(2\theta)\right] + \right.
\nonumber \\
&+& \left.  \sum_{\nu, f}\frac{\partial f} {\partial \epsilon}
\left[\epsilon_{\nu, f}(k')\right]\cdot \left[1\pm
\cos(2\theta)\right] \right\}
\end{eqnarray}
where we distinguish between \emph{interband} ($i\ne f$, first
line) and \emph{intraband} ($i=f$, second line) transitions, for
which we used the following limit:
\begin{equation}\label{DOS1}
\lim_{\mathbf{q}\rightarrow 0}\frac{f\left[\epsilon_{\nu,
f}(k')\right]-f\left[\epsilon_{\nu, f}(k'+q)\right]}
{\epsilon_{\nu, f}(k')-\epsilon_{\nu, f}(k'+q)} =\frac{\partial f}
{\partial \epsilon} \left[\epsilon_{\nu,
f}(k')\right].\end{equation} Then if the temperature $T\rightarrow
0$,
\begin{equation}\label{DOS2}
\frac{\partial f}{\partial \epsilon} \left[\epsilon_{\nu,
f}(k')\right]= -\frac{4\pi^2
d}{\sqrt{3}a_0^2}\sum_{\sigma=\pm}DOS_{\nu,
f,\sigma}(\epsilon_F)\delta(k'-k_{\nu, f,\sigma}^{F})
\end{equation}
where $DOS_{\nu, f,\sigma}(\epsilon_F)$ and $k_{\nu,
f,\sigma}^{F}$ are the electronic \emph{density of states} (DOS)
per $C$ atom and the Fermi wavevector of the $f$ band of $\nu$
index and slope of sign $\sigma$. The intraband EPC contribution
to the dynamical matrix at $\Gamma$ is thus proportional to the
density of states at the Fermi level.

The dynamical matrix at finite \textbf{q} can be calculated by
using the general expression~(\ref{eq:stat}); in the following we
will focus on the contribution at \textbf{q}=0 of the lowest
conduction band $\nu$=0, $\tilde
D_{\mathbf{\Gamma}}^{0}$,  and the second lowest bands $\nu=\pm
1$, $D_{\mathbf{\Gamma}}^{\pm 1}$, as functions of the Fermi energy
\begin{equation}\label{w_first_second}
\tilde D_{\mathbf{\Gamma}}(\epsilon_F)= \tilde
D_{\mathbf{\Gamma}}^{0}(\epsilon_F)+ \tilde
D_{\mathbf{\Gamma}}^{\pm 1}(\epsilon_F)
\end{equation}
The contribution of higher conduction bands could be in principle
included with no difficulty in the same form as the $\nu=\pm 1$
bands, but their effect is negligible for the doping levels
considered in this work, up to $N_{el}/C\sim 0.06$, $i.e.$ the
value of estimated total charge transfer in the case of alkali
intercalation~\cite{Lu04}.

\subsection{Contribution of the $\nu=0$ bands}

The bands intersecting at the \textbf{K} point of the graphene BZ
are at a good approximation linear around \textbf{K}, as shown in
the previous section, with slopes $\pm\beta= 14.1\cdot
\frac{a_0}{2\pi}$ $eV$. For these branches, $\nu=0$, so that the
geometric contribution $1\pm\cos(\theta+\theta')$ equals 0 or 2.
In practice, in the LO axial case it equals 2 when the transition
involves \emph{interband} states with opposite slope, and vanishes
for the \emph{intraband} transitions; the opposite behavior occurs in
the TO tangential case~\cite{Lazzeri05}.
For LO axial modes Eq.
\ref{eq:statgamma} becomes
\begin{equation}
\label{eq:first} \tilde
D_{\mathbf{\Gamma}}^{\mathrm{LO},0}=\frac{2 a_0^2 \sqrt{3}}{\pi^2
d} \langle G_\mathbf{\Gamma}^2
\rangle_\mathrm{F}\int_{-\overline{k}}^{\overline{k}} \mathrm{d}k'
\; \left[ \frac{f(\beta k') -f(-\beta k')}{\beta k'} \right]
\end{equation}
In the limit $T\rightarrow 0$, one obtains
\begin{equation}
\label{eq:statfirst} \tilde
D_{\mathbf{\Gamma}}^{\mathrm{LO},0}=\frac{4 a_0^2 \sqrt{3}}{\pi^2
d \beta} \langle G_\mathbf{\Gamma}^2 \rangle_\mathrm{F}
\ln\frac{|\epsilon_F|}{|\beta \overline{k}|}
\end{equation}
The quantitative result depends on the value of $\overline{k}$,
which enters the additive constant defined for $\omega_{LO}$ at
the end of the next subsection. Since the density of states is
independent of energy for the $\nu=0$ band, the non-analytical
direct EPC contribution to the TO tangential mode is also a
constant, independent of $\epsilon_F$
\begin{equation}
\label{eq:statfirst3} \tilde
D_{\mathbf{\Gamma}}^{\mathrm{TO},0}=-\frac{4 a_0^2 \sqrt{3}}{\pi^2
d \beta} \langle G_\mathbf{\Gamma}^2 \rangle_\mathrm{F}
\end{equation} which we include in the additive constant
defined for $\omega_{TO}$. In other words, our model predicts, in
the limit of low temperatures, a logarithmic divergence of the
dynamical matrix at $\Gamma$ for zero doping in the LO axial case,
while the TO tangential modes are simply red-shifted by the
non-analytical EPC contribution.

\subsection{Contribution of the $\nu=\pm 1$ bands}

Let us now consider the second lowest $\nu=\pm 1$ bands. Since
there is a large gap $E_{11}^M$ between the $\pi$ and $\pi^*$
bands, the interband transitions cannot give rise to KAs, and are
expected to contribute negligibly to the Fermi level dependence of
the phonon frequencies. On the other hand, in the case of doped
nanotubes where the $\nu=\pm 1$ band is partly populated,
\emph{intraband} transitions involve a density of states which
diverges in one dimension, leading to possible Kohn anomalies. We
develop the following analytical expressions based on the
ideal hyperbolic  $\nu=\pm 1$ bands.
According to Eqs. \ref{eq:statgamma},
\ref{DOS1}, and \ref{DOS2}, the contribution of intraband
transitions to $\tilde D_{\mathbf{\Gamma}}$ amounts to
\begin{equation}
\tilde D_{\mathbf{\Gamma}}^{\mathrm{TO/LO},1}=-8\langle
G_\mathbf{\Gamma}^2 \rangle_\mathrm{F} DOS_{ 1,\pi^*}(\epsilon_F)
\cdot [1\pm\cos(2\theta_F)] \label{eq:nextbands}
\end{equation}
where the $\nu\pm 1$ degeneracy is taken into account by a factor
of 2, and
\begin{equation}
DOS_{ 1,\pi^*}(\epsilon_F)=\theta(\epsilon_F - E_{11}^M/2) \frac{2
a_0^2 \sqrt{3}}{\pi \beta
d}\frac{\epsilon_F}{\sqrt{\epsilon_F^2-(E_{11}^M/2)^2}}\end{equation}
is the total DOS of the $\nu=1, \pi^*$ band (see
Eq.~\ref{fermieq}). The angular factor at the Fermi wavevector is
\begin{equation}
\cos(2\theta_F)=\frac{(k^F_{1,\pi^*})^2-k_\perp^2}{(k^F_{1,\pi^*})^2+k_\perp^2}
\end{equation}
where
$k^F_{1,\pi^*}=\pm\frac{1}{\beta}\sqrt{\epsilon_F^2-(E_{11}^M/2)^2}$
is the Fermi momentum of the $\nu=1, \pi^*$ band. We can notice
that the EPC direct contribution $\tilde
D_{\mathbf{\Gamma}}^{\mathrm{LO},1}$ is proportional to the
density of states of the $\nu=\pm 1$ conduction bands at the Fermi
energy, modulated by the cosine factor. Kohn anomalies would thus
occur \emph{where the density of states diverges}. With a few
algebraic calculations one obtains, for LO axial phonons, the
compact form:
\begin{equation}
\tilde D_{\mathbf{\Gamma}}^{\mathrm{LO},1}=-\frac{8 a_0^2
\sqrt{3}}{\pi^2 d \beta} \langle G_\mathbf{\Gamma}^2
\rangle_\mathrm{F} \frac{(E_{11}^M/2)^2}
{\epsilon_F\sqrt{\epsilon_F^2-(E_{11}^M/2)^2}}\cdot\theta(\epsilon_F-E_{11}^M/2)
\label{eq:statnextLO}
\end{equation}
For TO tangential modes through similar calculations one finally
obtains:
\begin{equation}
\tilde D_{\mathbf{\Gamma}}^{\mathrm{TO},1}=-\frac{8 a_0^2
\sqrt{3}} {\pi^2 d \beta} \langle G_\mathbf{\Gamma}^2
\rangle_\mathrm{F}
\frac{\sqrt{\epsilon_F^2-(E_{11}^M/2)^2}}{\epsilon_F}\cdot\theta(\epsilon_F-E_{11}^M/2)
\label{eq:statnextTO}
\end{equation}
In this latter case, the divergence of the DOS is canceled by the
cosine factor, which equals 0 for $\epsilon_F=E_{11}^M/2$. In the
following subsection, we will thus calculate through this model
the $\Gamma$ frequency of the TO/LO modes as:
\begin{equation}\label{omegasum}
\omega^\mathrm{TO/LO}_\mathbf{\Gamma}=
\sqrt{(\omega_{\mathbf{\Gamma}}^{\mathrm{TO/LO},\mathrm{other}})^2
+ \frac{\tilde D_{\mathbf{\Gamma}}^{\mathrm{TO/LO},0} + \tilde
D_{\mathbf{\Gamma}}^{\mathrm{TO/LO},1}}{M}},
\end{equation}
where
$\omega_{\mathbf{\Gamma}}^{\mathrm{TO/LO},\mathrm{other}}=\omega_{\mathbf{\Gamma}}+C^{\mathrm{TO/LO}}$
contains the unperturbed value
$\omega_{\mathbf{\Gamma}}=1581~cm^{-1}$, plus an additive constant
$C^{\mathrm{TO/LO}}$, which also includes the terms discussed at
the end of the previous subsection, and is chosen as to match the
DFT phonon-EZF results at zero doping; its numerical value, for
$\overline{k}=0.1 \frac{2\pi}{a_0}$; is 15.1/39.1 $cm^{-1}$ in the
TO/LO case respectively.

\subsection{Results: Kohn anomalies within the adiabatic approximations}
\begin{figure}
\centerline{\includegraphics[width=8.25cm]{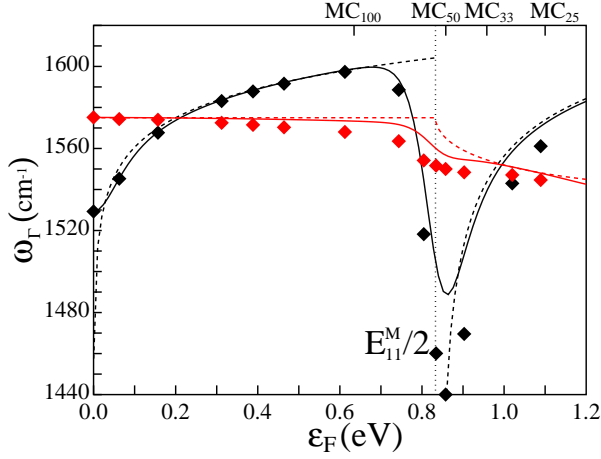}}
\caption{\label{adiabatic} (Color online) Variation, within the adiabatic
approximation, of the LO axial (black) and TO tangential (red)
$\Gamma$ frequencies with the Fermi level at T=315 K, calculated
from DFT phonon-EZF (full diamonds), from
the integral model (solid line, Eq.~\ref{eq:stat}), and from the
analytical model at T=0 K (dashed line, Eqs.~\ref{eq:statfirst},
\ref{eq:statnextLO}, and \ref{eq:statnextTO}). The metal/carbon
composition corresponding to the Fermi level is indicated on the
top of the figure.}
\end{figure}

At this point, the actual determination of the effect of EPC in
the vibrational properties of doped nanotubes can be in principle
carried out by adopting approximated calculation schemes. The DFT
phonon calculation of a SWNT is very demanding and is doable only
at a relatively high electronic temperature, where the potentially
interesting behaviors are smeared out. However, the comparison of
the calculations done on a real SWNT with those done using the
phonon-EZF (Fig.~\ref{compare}) shows that a quantitative
determination of the phonon frequency can be done neglecting the
nanotube curvature. Now, we will analize an integral model which
takes another step in identifying the key role played by
graphene-derived EPC in the vibrational properties of SWNTs, and
its low-temperature limit can be put into an analytical
expression. We report in Fig.~\ref{adiabatic} the frequencies of
the LO axial and the TO tangential modes of the $G$ peak at an
electronic temperature of 315 K as a function of the Fermi energy,
and thus of the electronic doping, as calculated $i)$ from the
phonon-EZF method; $ii)$ from the numerical integration of
Eq.~\ref{eq:stat} at T=315 K (polynomial fits are used to describe
the electronic bands); $iii)$ from its analytical limit at T=0 K,
where we use the ideal conical bands (Eqs.~\ref{eq:statfirst},
~\ref{eq:statfirst3},~\ref{eq:statnextLO},~\ref{eq:statnextTO},
and~\ref{omegasum}), with $E_{11}^M/2$ red-shifted by 0.06 $eV$ as
explained in subsection~\ref{subsectionFermi}.

A first inspection of the graphs indicates that the agreement
within the different approaches, and in particular between the DFT
calculations and the integral model, is excellent. The LO axial
frequency drops at $\epsilon_F=0$ due to a Kohn anomaly arising
from the first conduction band, and $\tilde
D_{\mathbf{\Gamma}}^{\mathrm{LO},0}$ varies as $\ln \epsilon_F$
according to Eq. \ref{eq:statfirst}. Another KA occurs when the
Fermi energy reaches the minimum of the second bands
$E_{11}^M/2=0.83$ $eV$.  This second KA is much stronger than that
at $\epsilon_F=0$, in agreement with Eq.
\ref{eq:statnextLO} which predicts a variation going as
$-1/\sqrt{\epsilon_F^2-(E_{11}^M/2)^2}$.
A frequency drop of about 40 $cm^{-1}$, with respect to zero doping,
was reported in a theoretical calculation of alkali-doped
SWNTs~\cite{Akdim05} at a doping level of about $N_{el}/C\sim
0.02$, in good agreement with our results (see
Fig.~\ref{compare}). Other KAs occur each time the Fermi level
increases such that a new conduction band is populated. In TO
tangential modes, due to the vanishing of the EPC interband matrix
element, the first band does not give rise to a KA; the curve is
practically flat until $E_{11}^M/2$. The effect of the second
conduction band induces a variation going as
-$\sqrt{\epsilon_F^2-(E_{11}^M/2)^2}$ (Eq.
\ref{eq:statnextTO}).

However, as we will see in the following section, although the
physics described by these results seems very intriguing, the main
outcome of this comparison is the accuracy of the integral model,
with respect to full DFT calculations, in predicting the Raman
properties of SWNTs, and in particular the anomalous doping
dependence of the vibrational axial mode.

\section{Non-adiabaticity}\label{nonadia}

\subsection{Time-dependent perturbation theory}

The vast majority, if not the totality, of \emph{ab initio}
phonon-calculations rely on the so-called Born-Oppenheimer adiabatic
approximation. In this framework one can decouple the electron
motion from the ion dynamics, on the basis of their large mass and
velocity difference, and thus treat the electronic properties as
they were completely independent of the ionic motion. This
approximation for phonon calculation is equivalent to first-order
time-{\it independent} perturbation theory and
is wholly justified for insulators and semiconductors.
Even in ordinary metals a proper,
specific treatment of the electronic degrees of freedom, such the
inclusion of a finite Fermi-Dirac electronic temperature, usually
suffices to avoid a failure of the BO approximation.

Metallic SWNTs do however represent an ``extraordinary'' case, due
to their intrinsic one-dimensional nature, and the use of the adiabatic BO
approximation leads to significant deviations from experimental
results. This could be particularly true in the case of Raman
scattering where a sinusoidal excitation induces the oscillatory
motion of ions. In order to check whether a proper inclusion of
non-adiabaticity would affect our previous results, we introduce a
non-adiabatic model, based on time-{\it dependent} perturbation theory.
In the process of absorption and emission of a phonon by the
electrons, the phonon energy is no longer neglected: we replace, in Eq. \ref{eq:stat},
the energy difference between the two electronic scattering states,$(\epsilon_{\mathbf{k},
i}-\epsilon_{\mathbf{k}+\mathbf{q}, f})$, by
$(\epsilon_{\mathbf{k},
i}-\epsilon_{\mathbf{k}+\mathbf{q}, f}+\hbar \omega_\mathbf{q}+i\delta)$.
Here we add a small
imaginary part $\delta$ to the energy
to control the divergences.
Eq. \ref{eq:stat} then becomes:
\begin{equation}\tilde D_{\mathbf{q}}^{\rm TO/LO}=\frac{a_0^2 \sqrt{3}\langle G_\mathbf{\Gamma}^2
\rangle_\mathrm{F}}{\pi^2 d} \sum_{\nu, f,
i}\int_{-\overline{k}}^{\overline{k}} \mathrm{d}k'
\;\frac{f\left[\epsilon_{\nu, i}(k')\right]- f\left[\epsilon_{\nu,
f}(k'+q)\right]}{\epsilon_{\nu, i}(k')-\epsilon_{\nu,
f}(k'+q)+\hbar\omega_\mathbf{q}+i\delta} \cdot \left[1\pm
sign(\epsilon_{\nu, f}\cdot\epsilon_{\nu,
i})\cos(\theta+\theta')\right] \label{eq:dyn}
\end{equation}
The frequency is obtained as $\omega_\mathbf{q}=
\sqrt{\frac{\Re\mathrm{e} (D_{\mathbf{q}})}{M}}$, that is, we take
the principal part of the above integral by letting
$\delta\rightarrow 0$. Since $D_{\mathbf{q}}$ depends on the frequency,
this equation should be solved
self-consistently.

\begin{figure}
\centerline{\includegraphics[width=8.25cm]{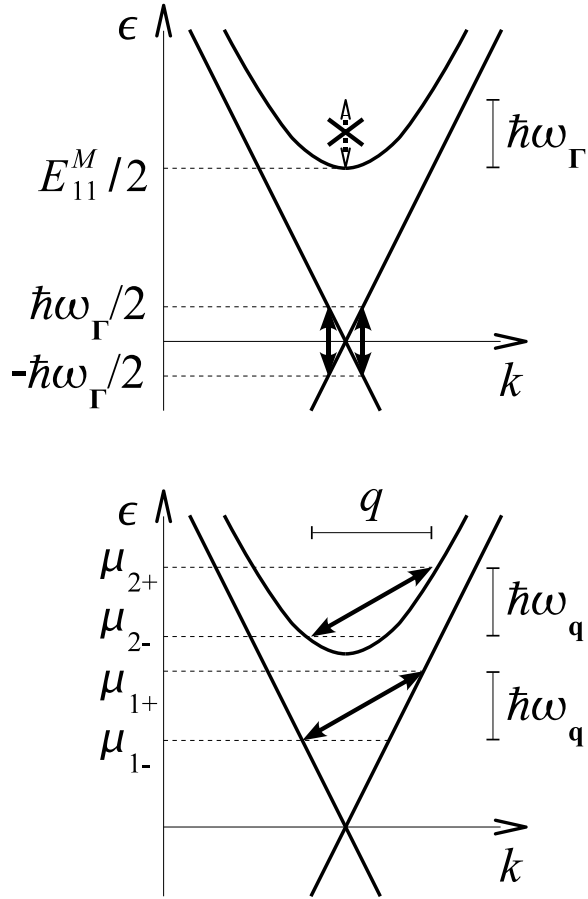}}
\caption{\label{transitions}. Schematic representation of
electronic transitions allowed by conservation of energy and
momentum in electron-phonon scattering processes involving a
phonon with momentum \textbf{q} and energy
$\hbar\omega_\mathbf{q}$. An abrupt change in the phonon
dispersion, {\it i.e.} a Kohn anomaly, occurs when, by change of
the Fermi energy, one of this transitions becomes allowed or
forbidden by the Pauli exclusion principle. In practice, this
occurs when the Fermi level crosses a tip of one of the arrows
corresponding to allowed transitions. Top panel: transitions at
\textbf{q}=\textbf{0} ($\mathbf\Gamma$). Bottom panel: transitions
at a finite \textbf{q}-point.}
\end{figure}
In the $\mathbf{q}\rightarrow \mathbf{0}$ limit,
Eq.~\ref{eq:statgamma} then becomes, in the NA case,
\begin{equation}\label{eq:dyngamma}
\tilde D_{\mathbf{\Gamma}}^{\rm TO/LO}= \frac{a_0^2
\sqrt{3}\langle G_\mathbf{\Gamma}^2 \rangle_\mathrm{F}}{\pi^2 d}
\int_{-\overline{k}}^{\overline{k}} \mathrm{d}k' \; \left\{
\sum_{\nu, f\ne i}\frac{f\left[\epsilon_{\nu, i}(k')\right]-
f\left[\epsilon_{\nu, f}(k')\right]}{\epsilon_{\nu,
i}(k')-\epsilon_{\nu, f}(k')+\hbar\omega_\mathbf{\Gamma}+i\delta}
\cdot \left[1\pm sign(\epsilon_{\nu, f}\cdot\epsilon_{\nu,
i})\cos(2\theta)\right]\right\}
\end{equation}
Here only the interband transitions contribute to the
dynamical matrix, since, for an optical phonon, the limit
\begin{equation}\label{DOS3}
\lim_{\mathbf{q}\rightarrow 0}\frac{f\left[\epsilon_{\nu,
f}(k')\right]-f\left[\epsilon_{\nu, f}(k'+q)\right]}
{\epsilon_{\nu, f}(k')-\epsilon_{\nu,
f}(k'+q)+\hbar\omega_\mathbf{q}+i\delta} =0
\end{equation}
We can anticipate that the absence of the intraband contributions
dramatically affects the results when such terms are important,
\emph{i.e.} when the Fermi level is close to a $E_{\nu\nu}^M/2$
band minimum. As in the previous section, we study the
contribution of the $\nu=0,\pm 1$ bands to the non-analytical part
of the dynamical matrix. The mode frequencies are obtained from
the dynamical matrix as in Eq.~\ref{omegasum}.

\subsection{Contribution of the $\nu=0$ bands}

Eq.~\ref{eq:dyngamma} can be integrated, by using linear bands and
$T=0$, analogously to the calculations developed in the static
case; we thus obtain for the non-analytical part of
$D_\mathbf{\Gamma}$ in the case of LO axial phonons
\begin{equation}
\tilde D_{\mathbf{\Gamma}}^{\mathrm{LO},0}=\frac{2 a_0^2
\sqrt{3}}{\pi^2 d \beta} \langle G_\mathbf{\Gamma}^2
\rangle_\mathrm{F}
\ln\frac{|2\epsilon_F+\hbar\omega^\mathrm{LO}_\mathbf{\Gamma}||2\epsilon_F-
\hbar\omega^\mathrm{LO}_\mathbf{\Gamma}|}{|2\beta \overline{k}|^2}
\label{eq:dynfirst}
\end{equation}
while the TO tangential term $\tilde
D_{\mathbf{\Gamma}}^{\mathrm{TO},0}$ is zero since it involves the
intraband terms only. We note that the KA observed in the static
case for zero doping is replaced by two logarithmic divergences at
a doping level $\epsilon_F=\pm\hbar \omega^\mathbf{LO}_\mathbf{\Gamma}/2$, which
is $N_{el}/C\approx 0.0015$ ($\sim$ MC$_{\rm 650}$). As shown
schematically in the top panel of Fig.~\ref{transitions}, $\Gamma$
transitions conserving energy and momentum in the scattering
process can indeed occur when connecting two bands separated by an
energy equal to $\hbar\omega_\mathbf{\Gamma}$, which is in
practice the case only in correspondence of the electronic
vertical transitions (\textbf{q}=\textbf{0}) between the $\pi$ and
the $\pi^*$ bands.

\subsection{Contribution of the $|\nu|\ge1$ bands}
The effect of the non-adiabatic terms in the $\nu=0$ bands
contribution merely consists in a splitting of the zero-doping
Kohn anomaly observed in the static case. The picture changes
completely for the $|\nu|\ge1$ bands. In fact, given the energy
gap between the $\pi$ and $\pi^*$ bands having $\nu\neq 0$, only
the intraband terms are relevant to Kohn anomalies. These intraband
terms are suppressed by the dynamical effects
(Eq.~\ref{eq:dyngamma}) in both LO and TO modes and thus $\tilde
D_{\mathbf{\Gamma}}^{\mathrm{LO,TO},\nu\neq 0}=0$.
Indeed in the hyperbolic band it is not possible to conserve
both energy and momentum in an electron-phonon scattering process
with a ${\rm q}=0$ optical phonon, see Fig.~\ref{transitions}.
In addition to the analytical calculation at T=0, we performed
numerical integrations of Eq.~\ref{eq:dyngamma} at different
temperatures (77 K, 315 K). As in the static case, we used
polynomial fits to describe the band dispersions. We report in
Fig.~\ref{nonadiaGamma} the variation of the LO axial mode as
function of the Fermi level in the non-adiabatic case at those two
temperatures. The results are consistent with the analytical ones
obtained at vanishing temperature; the only Kohn anomaly is
observed, as explained in the previous subsection, at
$\hbar\omega_\mathbf{\Gamma}/2$. All the other Kohn anomalies,
that in the adiabatic approximation and thus in all standard DFT
calculations are due to the population of new energy levels,
completely disappear. This is a general behavior for vertical
transitions.

\subsection{LO phonon linewidth}

\begin{figure}
\centerline{\includegraphics[width=8.25cm]{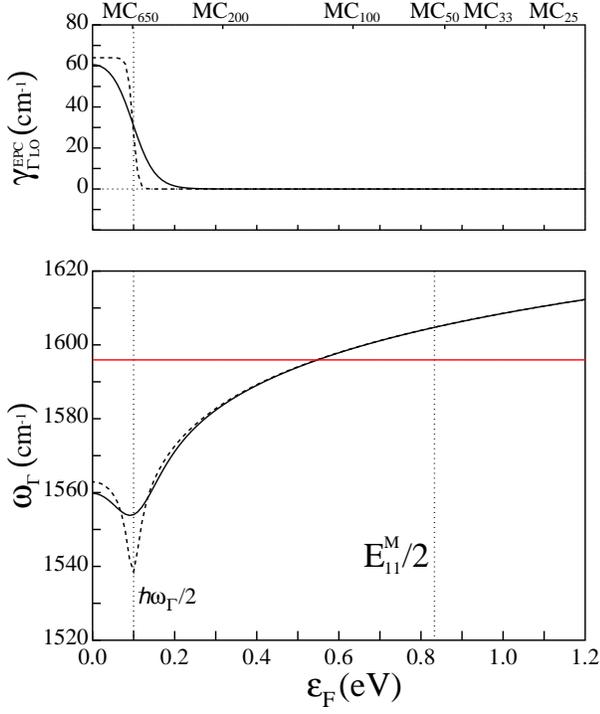}}
\caption{\label{nonadiaGamma}
(Color online) Lower panel: variation of the LO
axial ($G^-$ peak in Raman experiments, black) and TO tangential
($G^+$ peak, red) $\Gamma$ frequencies with the Fermi level at
T=315 K (solid line) and at 77 K (LO axial, dashed line),
calculated from the \textbf{non-adiabatic} integral model
(Eq.~\ref{eq:dyn}). Top panel: variation, at the same
temperatures, of the EPC contribution to the FWHM of the LO axial
mode with the doping level. The metal/carbon composition
corresponding to the Fermi level is indicated on the top of the
figure.}
\end{figure}

An important measurable quantity is the phonon linewidth. In
ref.~\cite{Lazzeri06} it is shown that it can be split into a EPC
direct contribution, only relevant in metallic nanotubes, and a
term due to inhomogeneous broadening and anharmonicites, common to
all nanotubes and estimated around 10$~cm^{-1}$ from experimental
data. The LO axial phonon linewidth (full width at half maximum)
$\gamma^{EPC}_{LO}$ as function of the electronic doping can
instead be calculated~\cite{JonesMarch} from the imaginary part of
the non-adiabatic dynamical matrix, Eq.~\ref{eq:dyn}, as
$\gamma_{\mathbf{q}}=|\Im(\tilde
D_{\mathbf{q}})|/(\omega_{\mathbf{q}} M)$. An identical result is
obtained by using the Fermi golden rule, as in
ref.~\cite{Lazzeri06}. The result of ref.~\cite{Lazzeri06} derived
in nanotubes for zero doping can be easily generalized for any
doping as:
\begin{eqnarray}
\gamma^\mathrm{EPC}_\mathrm{LO}(\epsilon_F) &=& \frac{2 a_0^2
\sqrt{3}}{\pi d \beta  M \omega^\mathrm{LO}_\mathbf{\Gamma}} \langle
G_\mathbf{\Gamma}^2
\rangle_\mathrm{F}\left[\frac{1}{e^{(\epsilon_F-\hbar\omega^\mathrm{LO}_\mathbf{\Gamma}/2)/k_B
T}+1}-\frac{1}{e^{(\epsilon_F+\hbar\omega^\mathrm{LO}_\mathbf{\Gamma}/2)/k_B
T}+1}\right] \\ \nonumber &=& \frac{79~cm^{-1} nm}{d}
\left[\frac{1}{e^{(\epsilon_F-\hbar\omega^\mathrm{LO}_\mathbf{\Gamma}/2)/k_B
T}+1}-\frac{1}{e^{(\epsilon_F+\hbar\omega^\mathrm{LO}_\mathbf{\Gamma}/2)/k_B
T}+1}\right]
\end{eqnarray}
At zero doping this FWHM equals, at T=315 K, about 60 $cm^{-1}$
for a (9,9) SWNT. We show the EPC LO linewidth in the top panel of
Fig.~\ref{nonadiaGamma}. At $T\rightarrow 0$ this term is constant
up to the Kohn anomaly at a $N_{el}/C\approx 0.0015$ (MC$_{\rm
650}$) doping level, and vanishes abruptly for higher doping
levels, while a smoother behavior is observed at finite
temperature. Since all the terms contributing to the EPC part of
the dynamical matrix vanish for the TO tangential modes, the
corresponding phonon linewidth will vanish as well.

\subsection{Finite-q results}

\begin{figure}
\centerline{\includegraphics[width=8.25cm]{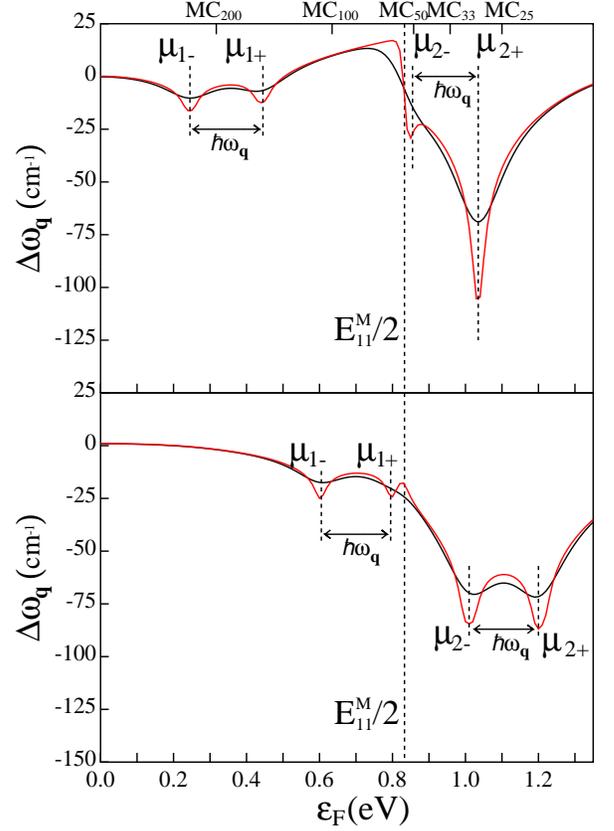}}
\caption{\label{nonadia_q}
(Color online) Lower panel: variation, with respect
to zero doping, of the LO axial frequencies with the Fermi level
at T=315 K (black) and T=77 K (red), calculated from the
\textbf{non-adiabatic} integral model (Eq.~\ref{eq:dyn}). The
phonon momentum is equal to \textbf{q}=0.1 $\cdot
\frac{2\pi}{a_0}$. Top panel: same, at \textbf{q}=0.05 $\cdot
\frac{2\pi}{a_0}$. The dotted vertical line indicates the minimum
of the second conduction band. The energy values $\mu_{1-}$,
$\mu_{1+}$, $\mu_{2-}$, and $\mu_{2+}$ correspond to the levels
schematically shown in Fig.~\ref{transitions}. A similar behavior
is expected for the highest optical phonon near the \textbf{K}
point, which is responsible for the Raman $D$ band. The
metal/carbon composition corresponding to the Fermi level is
indicated on the top of the figure.}
\end{figure}

In the previous subsections, we have developed our analytical
model at the $\Gamma$ point of the Brillouin zone, and we have
shown that the proper treatment of non-adiabatic terms lifts all
the divergences resulting from intraband, vertical transitions.
However, as shown schematically in the lower panel of
Fig.~\ref{transitions}, at finite \textbf{q} the outcome is
different from the $\Gamma$ case. Indeed, non-vertical intraband
transitions conserving energy and momentum can occur when
\textbf{q} is approximately larger than $\omega_q/\mathrm{v_{M}}$,
$\mathrm{v_{M}}$ being the maximum electronic band velocity. We
report in Fig.~\ref{nonadia_q} the dependence on electronic doping
of the LO axial frequency at \textbf{q}=$0.05 \cdot
\frac{2\pi}{a_0}$ and $0.1 \cdot \frac{2\pi}{a_0}$, calculated
numerically through Eq.~\ref{eq:dyn} at T=77 K and 315 K. In this
case the intraband Kohn anomalies observed in the adiabatic
approximation are not suppressed, as it is the case for vertical
transitions. The interband KA is shifted towards higher doping
levels with respect to the $\Gamma$ case. The intraband KAs are
shifted in the same direction with respect to the energy levels
$E_{\nu\nu}^M/2$. Both interband and intraband Kohn anomalies
appear in doublets, separated by $\hbar \omega_\mathbf{q}$.

By using the expressions for EPC derived in
Refs.~\cite{Piscanec04,Lazzeri05} for the scattering involving
phonons near the \textbf{K} point, it can be shown that the
behavior of the highest optical phonon at a wavevector
\textbf{q}+\textbf{K} is very similar to the one reported in
Fig.~\ref{nonadia_q} for the LO axial frequency at \textbf{q}.
This phonon near the \textbf{K} point is responsible for the $D$
band observed experimentally in Raman spectra.

\section{Discussion}\label{conclusions}

\subsection{Physical properties of metallic nanotubes}

In this work we presented a combination of \emph{ab initio}
DFT calculations with an integral and analytical model developed
to take into account in a simple way the direct EPC contribution
to the dynamical matrix, responsible for the divergences
originating the KAs. Since however the mathematics behind our
models might be cumbersome, here we
summarize the most interesting \emph{physical} properties of
metallic SWNTs we observe, as well as the quantitative results on
experimentally measurable quantities.

We showed that KAs in SWNTs can be identified by considering
electron-phonon scattering processes that conserve both energy and
momentum (see Fig.~\ref{transitions} and relative discussion).
Since there are strong disagreements between the adiabatic and the
non-adiabatic case, only the results obtained through the latter
model are reliable. The failure of the adiabatic calculation
originates from the fact that, in this approximation, the phonon energy is
neglected and the energy conservation in the electron-phonon
scattering processes is thus violated. Since the curves shown in
Figs.~\ref{compare} and~\ref{adiabatic} are obtained through the
adiabatic model and through DFT frozen-phonon and DFPT
calculations, those results should not be used for comparison to
experimental data. On the other hand, all the results presented in
Figs.~\ref{nonadiaGamma} and ~\ref{nonadia_q}, which conflict with
any existing DFT calculation based on the BO approximation, are in
principle \emph{correct}, and are thus meant to be directly
compared to experiments.

The results shown in Fig.~\ref{nonadiaGamma} promise to be useful
tools to determine experimentally the doping level of individual
metallic SWNTs. In particular, by measuring the shift of the $G^-$
peak with respect to zero doping and its linewidth, one can
determine the variation of the Fermi energy and thus the doping
level. Those results refer to a (9,9) nanotube of diameter 1.23
$nm$; however, they can be easily generalized to any $d$-diameter
metallic SWNT by observing that the dependence of both the
linewidth of the $G^-$ peak and its frequency shift with respect
to zero doping case are proportional to $1/d$ and independent of
the chirality. We note here that we are assuming that the $G$ peak
originates from single resonance.

It was showed that carbon nanotubes~\cite{Lu04,Margine06}, as
graphite~\cite{Posternak83,Csanyi05}, present a free-electron-like
interlayer state whose energy level decreases rapidly with
increasing doping, and reaches the Fermi energy at about the
saturation doping of $\sim$MC$_{\rm 8}$. We  showed that KAs
occur at $\mathbf\Gamma$ only when the Fermi energy is in
proximity of a band crossing with non-zero interband
electron-phonon coupling. Besides the crossing of the $\pi$-bands
at zero doping, we
estimate that the first band-crossing with symmetry-allowed
transitions is that between the interlayer level and the $\pi^*$
($\nu=0$) bands. Such crossing is a candidate for a Kohn anomaly.
This could explain the strong softening of about $50~cm^{-1}$
observed at saturation doping in experimental Raman studies on the
$G$ peak in alkali-doped SWNTs~\cite{Bendiab01,Sauvajol03}.

Finally, we predicted KAs to occur in phonons with \emph{finite}
momentum \textbf{q} not only in proximity of a band crossing, but
also each time a new band is populated (see Fig.~\ref{nonadia_q}).
Therefore, by experimentally probing such KAs, the filling of the
hyperbolic bands could be detected. Phonons at finite-\textbf{q}
vectors are experimentally accessible to double-resonant Raman
scattering~\cite{Thomsen00}, and correspond in the Raman spectra,
\emph{e.g.}, to the defect-activated $D$ peak at about 1350
$cm^{-1}$, and to the second-order (two-phonon) $2D$ and $2G$
peaks at approximately 2700 and 3200 $cm^{-1}$. We showed in
Fig.~\ref{nonadia_q} the variation of the phonon frequency with
respect to doping at \textbf{q}=$0.05 \cdot \frac{2\pi}{a_0}$ and
$0.1 \cdot \frac{2\pi}{a_0}$. These phonon momenta are
experimentally accessible to double-resonant Raman
scattering~\cite{Maultzsch03}.

Although our computational study has been limited to a metallic
SWNT, our analytical results allow to predict the behavior of the
high-energy Raman peaks in semiconducting nanotubes. Since in this
case only intraband terms contribute to the EPC term of the
dynamical matrix, we expect the $G$ peak to be unaffected by
electron doping. On the other hand, analogously to the metallic
case, the defect-activated $D$ peak should be sensitive to doping,
provided that the excitation energy is sufficient to populate the
first excited electronic level in the conduction band.

\subsection{Conclusions}

In this work we presented a theoretical study on the
vibrational/Raman properties of electron-doped SWNTs using
Density-Functional Theory and analytical models. We performed our
calculations within the adiabatic Born-Oppenheimer approximation,
but we also used time-dependent perturbation theory to explore
non-adiabatic effects beyond this approximation. We showed that
the Born-Oppenheimer approximation, predicts, for increasing
doping levels, a series of EPC-induced KAs in the vibrational mode
parallel to the tube axis at the $\mathbf\Gamma$ point of the
Brillouin zone, usually indicated in Raman spectroscopy as the
$G^-$ peak. Such Kohn anomalies would arise each time a new
conduction band is populated. However, we showed that they are an
artifact of the adiabatic approximation, which is the standard
approach for ab-initio phonon calculations. The inclusion of
non-adiabatic effects dramatically affects the results, predicting
KAs at $\mathbf\Gamma$ only when $\epsilon_F$ is close to a band
crossing $E_{X}$. For each band crossing a double KA occurs for
$\epsilon_F=E_{X}\pm \hbar\omega/2$, where $\hbar\omega$ is the
phonon energy. In particular, for a 1.2 $nm$ metallic nanotube, we
predicted a KA to occur in the so-called $G^-$ peak at a doping
level of about $N_{el}/C=\pm 0.0015$ atom ($\epsilon_F\approx \pm
0.1 ~eV$) and, possibly, close to the saturation doping level
($N_{el}$/$C$$\sim$ 0.125), where an interlayer band crosses the
$\pi^*$ nanotube bands. Furthermore, we predicted that the Raman
linewidth of the $G^-$ peak significantly decreases for
$|\epsilon_F| \geq \hbar\omega/2$. Thus our results provide a tool
to determine experimentally the doping level from the value of the
frequency shift and from the linewidth of the $G^-$ peak. Finally,
we predicted Kohn anomalies to occur in phonons with finite
momentum \textbf{q} not only in proximity of a band crossing, but
also each time a new band is populated. Such Kohn anomalies should
be observable in the double-resonant Raman peaks, such as the
defect-activated $D$ and peak, and the second-order peaks $2D$ and
$2G$. We also predict that in semiconducting nanotubes the $G$
peak should be insensitive to doping, while the $D$ peak should be
affected analogously to the metallic case.

We thank N. Bendiab, S. Piscanec and A. C. Ferrari for useful
discussions. DFT calculations have been performed at the IDRIS
French National Computational Facility under the projet CP9-61387.

\end{document}